\documentclass[aps,prb,twocolumn,showpacs]{revtex4}
\usepackage{amssymb}
\usepackage{amsmath}
\usepackage{graphicx}

\begin{document}

\title{Strong-coupling superconductivity revealed by scanning tunneling microscope in tetragonal FeS}

\author{Xiong Yang, Zengyi Du, Guan Du, Qiangqiang Gu, Hai Lin, Delong Fang, Huan Yang, Xiyu Zhu, and Hai-Hu Wen$^{\star}$}

\affiliation{National Laboratory of Solid State Microstructures and Department of Physics,
Collaborative Innovation Center of Advanced Microstructures, Nanjing University, Nanjing 210093, China}

\begin{abstract}
We investigate the electronic properties of the tetragonal FeS superconductor by using scanning tunneling microscopy/spectroscopy. It is found that the typical tunneling spectrum on the top layer of sulfur can be nicely fitted with an anisotropic \emph{s} wave or a combination of two superconducting components in which one may have a highly anisotropic or nodal-like superconducting gap. The fittings lead to the superconducting gap of about $\Delta_{max}\approx$ 0.90$\;$meV, which yields a ratio of 2$\Delta_{max}/k_BT_c\approx$ 4.65. This value is larger than that of the predicted value 3.53 by the BCS theory in the weak-coupling limit, indicating a strong-coupling superconductivity. Two kinds of defects are observed on the surface, which can be assigned to the defects on the S sites (fourfold image) and Fe sites (dumbbell shape). Impurity-induced resonance states are found only for the defects on the S sites and stay at zero-bias energy.
\end{abstract}


\maketitle

\section{INTRODUCTION}

The discovery of superconductivity in LaFeAsO$_{1-x}$F$_x$ with $T_c=26\;$K in early 2008 marked the beginning of worldwide investigation on this new family of unconventional superconductors\cite{Kamihara2008}. This also provides a good opportunity for investigating high-$T_c$ superconductivity\cite{Chubukov2011} in pursuing a generic picture. In iron-based superconducting families, so far superconductivity has been found in the iron pnictides (FeAs and FeP based) and chalcogenides (FeSe based)\cite{Stewart2011,Greene2010,Johnston2010}. Similar to cuprates, iron-based superconductors exhibit a layered structure with superconducting planes separated by charge tuning layers, and the superconductivity can be achieved by hole or electron doping or using chemical or high pressure when the antiferromagnetism of the parent state is suppressed\cite{WenHH2011}. In the study on iron-based superconductors, one of the core issues is the pairing mechanism which remains unresolved. It is proposed that the pairing might be mediated by the antiferromagnetic spin fluctuations with an $s^\pm$ wave\cite{Mazin2008,Kuroki2008,Hirschfeld2011}, which is satisfied in many FeAs- and FeSe-based systems with both hole and electron pockets and has several experimental confirmations\cite{Neurton2008,Hanaguri2010,YangH2013}. FeSe has the simplest atomic structure among all iron-based superconductors\cite{WuMK2008}. Nodal superconducting gaps and an electronic nematic state have been suggested\cite{XueQK2011} in FeSe thick films or bulk. Interestingly, the mono-layer FeSe film grown on SrTiO$_3$ substrate has an onset superconducting transition temperature $T_c$ reaching about 50 K, as indicated by the direct transport and magnetic measurements\cite{MaXC2014}. However, differently from bulk FeSe, there are no hole pockets, gap nodes, or electronic nematic phase observed in mono-layer FeSe thin film. Recently, as an isostructure of the FeSe superconductor, superconductivity has been reported with $T_c=4.5\;$K\cite{HuangFQ2015} in tetragonal FeS. Previous band structure calculations\cite{DuMH2008} have shown that the electronic structure of FeS is quite similar to FeSe. Therefore, it is intriguing to know whether the gap structure and the pairing mechanism of this newly discovered superconductor are similar to that of FeSe systems.

Superconductivity emerges when electron pairs with opposite momenta condense into a coherent state, and the superconducting gap protects the condensate from exciting quasiparticles. In the study of superconductivity, atomic defects as phase indicators can be used to measure the superconducting gap structure\cite{Hoffman2013}. According to Anderson's theorem\cite{Anderson1959,ZhuJX2006}, Cooper pairs with singlet pairing can survive in the presence of nonmagnetic defects, whereas the magnetic impurities are detrimental to superconductivity. This phenomena has been nicely demonstrated in conventional superconductors\cite{Yazdani1997} using Mn as the magnetic impurities. In contrast to such sign reserved gap, in the scenario with sign change or reversal gap symmetry, impurity effects will mix gaps on different parts of the Fermi surface and thereby smear out the momentum dependence\cite{Mishra2009}. In this way, both scalar potential and magnetic impurities can induce in-gap states in superconductors with $d$-wave and sign-reversal $s$-wave gaps\cite{PanSH2000,Hudson2001,YangH2013}. Scanning tunneling microscopy/spectroscopy (STM/STS) is a direct probe to detect the local density of states (LDOS), which can provide key information on the superconducting gap symmetry\cite{Hoffman2011}. With the ability of high spatial and energy resolution, STM/STS has proven to give significant impact in detecting the gap structure by observing the characterization of density of states near the defects\cite{PanSH2000,Hudson2001,Hudson2008,XueQK2011,Hoffman2011,YangH2013,Pennec2012,PanSH2015}.

In this paper, we report the first STM/STS experiment of the newly discovered iron-based superconductor FeS. The superconducting tunneling spectrum can be fitted well with the anisotropic $s$ wave or $s+d$ waves by using Dynes model. The corresponding gap ratio of the 2$\Delta/k_BT_c$ is larger than the predicted one by the BCS theory in the weak coupling limit, suggesting strong coupling in this compound. We observed two kinds of defects on the surface. The defect located at Fe site gives negligible influences on the STS spectra and the superconducting gap, while the defect located at S site induces clear in-gap states with the peak locating at zero energy.

\begin{figure}
\includegraphics[width=8.5cm]{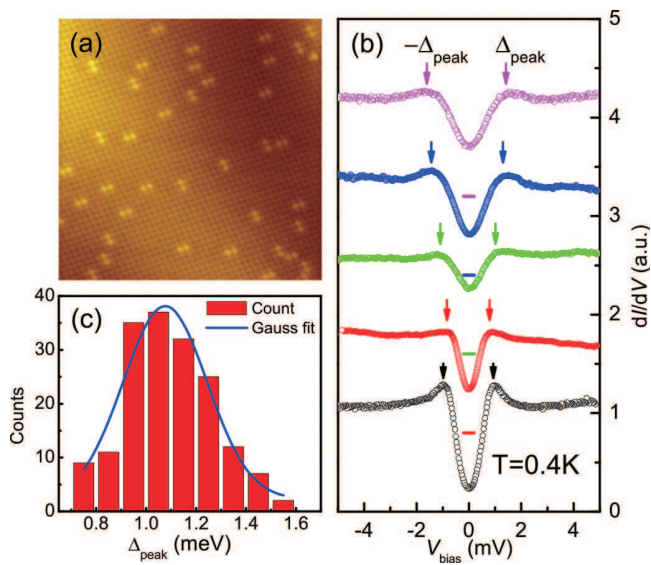}
\caption {(color online) (a) High-resolution topographic STM image (15 $\times$ 15 nm$^2$) of the cleaved surface constituted by S atoms. (b) Five representative STS spectra with different coherence-peak energies measured at 0.4$\;$K. The STS spectra in panel (b) are offset for clarity. The colored horizontal bars indicate the zero-conductance position for each corresponding spectrum, and colored arrows indicate coherence-peak energies. (c) Distribution of the coherence-peak energies based on statistics of about 170 STS spectra measured in different areas of the sample at 0.4$\;$K. The solid blue line shows a Gaussian fit to the distribution with the mean value of 1.08$\;$meV.}
\label{fig1}
\end{figure}

\section{EXPERIMENTAL METHOD}

The FeS crystals with tetragonal structure were grown by a hydrothermal method which has been presented previously\cite{LinH2016}. The STM/STS was measured with an ultrahigh vacuum, low temperature, and high magnetic field scanning probe microscope, USM-1300 (Unisoku Co., Ltd.). The FeS crystal was cleaved at room temperature in a high vacuum chamber with pressure better than 10$^{-10}$$\;$torr and then quickly transferred into the microscope head, which was kept at a low temperature. Measurements were conducted on the freshly cleaved surface of the sample. The electrochemically etched tungsten tips were treated by \emph{in situ} electron-beam sputtering and then used in all STM/STS measurements. A lock-in amplifier with modulation frequency $0.3\;$mV at $987.5\;$Hz was used to lower the noise of the differential conductance spectra.

\section{RESULTS AND DISCUSSION}

Fig.~\ref{fig1}(a) shows an atomically resolved topographic image of the terminated surface of cleaved FeS single crystal. The lattice constants in the perpendicular square directions are 3.69 and 3.62 \AA, respectively, and the value is comparable to 3.68 \AA from previous report\cite{HuangFQ2015}. The slight lattice difference between two perpendicular directions may come from the system error of STM. Since the lattice constant is very close to the expected distance of an S-S bond and the S-atom layer is the natural termination surface, we can reasonably conclude that the terminated top layer is the S-atom layer. There are some dumbbell-shaped impurities on the surface, which are very similar to that of Cu impurity from our previous work\cite{YangH2013}. The impurity atom in the center of dumbbell spot locates between two S atoms in the terminated top layer and at the position of Fe atom in the layer beneath. These impurities may be induced by some vacancies at the Fe sites or the partial substitution of Fe atoms with S or K atoms during the crystal growth process\cite{Vacancy2012,XueQK2012}. The STS spectra measured on the surface shown in Fig.~\ref{fig1}(a) exhibit homogeneous features, while at different locations of the sample, the STS spectra may show different coherence-peak energy and zero-bias conductance. Figure~\ref{fig1}(b) shows five typical STS spectra with different coherence-peak energy measured in different areas of the sample. The statistics on the coherence-peak energy of 170 STS spectra with superconducting feature measured in different areas of the sample, displayed in Fig.~\ref{fig1}(c), show clearly that the coherence-peak energy locates between 0.7 and 1.6 meV. A mean value of the coherence-peak energy of about 1.08 $\pm$ 0.17 meV is obtained by Gauss fitting. The defect density in different areas where the STS spectra are measured are different, which may cause the wide distribution of the coherence-peak energies.

The experimental tunneling curve \emph{dI/dV} vs. \emph{V} can be fitted by the Dynes model\cite{Dynes1978} with tunneling current for one gap of
\begin{equation}
\begin{aligned}
I(V)\propto&\int_{-\infty} ^\infty \mathrm{d}\varepsilon \int_0 ^{2\pi}\mathrm{d}\theta[f(\varepsilon)-f(\varepsilon+eV)]\\
&\times\mathrm{Re}\left\{\frac{\varepsilon+eV-i\Gamma}{[(\varepsilon+eV-i\Gamma)^2-\Delta^2(\theta)]^{1/2}}\right\},
\end{aligned}
\end{equation}
where $\Gamma$ is the broadening parameter, $\Delta(\theta)$ is the superconducting gap function, and $f(\varepsilon)$ is the Fermi distribution function containing the thermal broadening effect at some finite temperature. Both the temperature and the broadening parameter $\Gamma$ will influence coherence peaks and zero-bias conductance, causing ungapped DOS within superconducing gap. The STS spectra in Fig.~\ref{fig1}(b) are measured at 0.4$\;$K and show finite zero-bias conductance. At 0.4$\;$K, the influence of temperature will be small, indicating that the finite zero-bias conductance is mainly caused by the large broadening parameter $\Gamma$. The FeS sample we used here has a residual resistivity of about 60 $\mu\Omega$ cm\cite{LinH2016}, which is much larger than the one about 10 $\mu\Omega$ cm of FeSe\cite{Mastuda2014}. This may cause large broadening parameter $\Gamma$, giving a finite zero-bias conductance. With large broadening parameter $\Gamma$, no matter what superconducting gap function is, the coherence-peak energy value $\Delta_{peak}$ will always be larger than the real superconducting gap $\Delta$. Thus, in order to verify the pairing symmetry of FeS clearly, we should choose the spectrum with low zero-bias conductance to do further analysis.

\begin{figure}
\includegraphics[width=8.5cm]{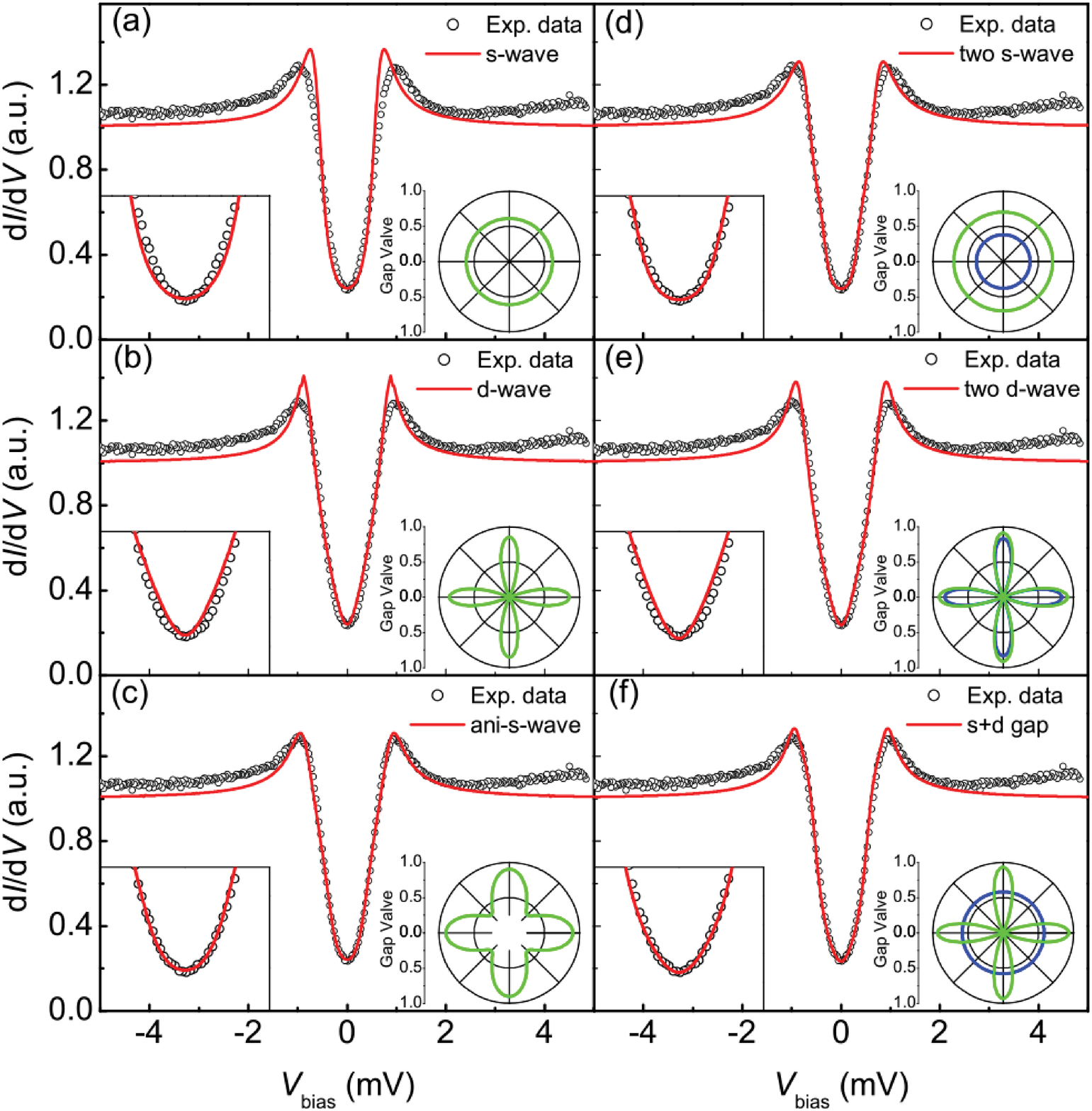}
\caption {(color online) The Dynes model fitting to an STS spectrum measured at 0.4$\;$K. The open circles represent the experimental data, and the red lines are the theoretical fits to the data with the Dynes model by (a) single \emph{s} wave, (b) single \emph{d} wave, (c) single anisotropic \emph{s} wave, (d) double \emph{s} waves, (e) double \emph{d} waves, (f) \emph{s} + \emph{d} waves. The left-hand-side inset in each figure shows the closeup of the corresponding fitting results. The right-hand-side inset shows the gap function for each case. The green and blue curves in the right inset for each case stand for the component(s) to construct the gap function.}
\label{fig2}
\end{figure}

Therefore, we take the spectrum with well-resolved coherence peaks and low zero-bias conductance, shown in Fig.~\ref{fig1}(b) as black symbols, to do the fitting. The fitting results with several scenarios of superconducting gap functions based on the Dynes model are shown in Fig.~\ref{fig2}. The gap functions used in the fitting are formed by one or two component(s) based on isotropic \emph{s}-wave, anisotropic \emph{s}-wave, or \emph{d}-wave gaps, respectively. For purposes of inspecting the fitting results more clearly, we zoom in the bottom of the spectrum together with the fitting curve as shown in the left inset of each figure. For the \emph{s}-wave fitting, as shown in Fig.~\ref{fig2}(a), the calculation which displays a more flat bottom compared with the experimental data fails to track the low energy line shape. On the other hand, the fitting with a single \emph{d}-wave gap shown in Fig.~\ref{fig2}(b) generates a more V-shaped feature near the bottom, which also deviates from the experimental data. Both the coherence peaks of the two fitting results are much sharper than the experimental data, so we use an anisotropic \emph{s}-wave to simulate the data. As shown in Fig.~\ref{fig2}(c), the anisotropic \emph{s} wave can fit the data well with the gap function of $\Delta$($\theta$)=$\Delta$(0.7+0.3cos4$\theta$). Considering the multiband features in this system\cite{LinH2016}, we also used two components ($s_1 + s_2$ waves, $d_1 + d_2$ waves, or $s + d$ waves) with each containing a single gap function (either $s$ or $d$ wave). Among these fitting results with two wave functions, the simulation with $s + d$ waves shown in Fig.~\ref{fig2}(f) can nicely fit the data. All the detailed fitting parameters for different models are shown in Table 1. With the status of our experimental data, we cannot conclude which one of the anisotropic \emph{s} wave or the $s + d$ wave gap function is better. So we argue that the superconducting gap is highly anisotropic, or even has nodes\cite{XingJ2016,WangQH2016}. Besides, both fittings of the anisotropic \emph{s} wave and the $s + d$ wave gap function lead to the similar maximum superconducting gap $\Delta_{max}\approx$ 0.90$\;$meV, which yields a ratio of 2$\Delta_{max}/k_BT_c\approx$ 4.65. This value is larger than that of the predicted value 3.53 by the BCS theory in the weak coupling limit, indicating a strong-coupling superconductivity in this system.

\begin{table}
\caption{Fitting parameters with different models for FeS. The units of $\Delta$ and $\Gamma$ are meV. }
\begin{tabular}
{ccccccc}\hline \hline
Model & $\Delta_1$ &$\Gamma_1$ &\emph{p} &$\Delta_2$ &$\Gamma_1$ &1-\emph{p} \\
\hline
$s \quad wave$  & 0.61 & 0.15  & 100\%    &  -  &  -   &  - \\
$d \quad wave$  & 0.85 & 0.08  & 100\%    &  -  &  -   &  - \\
$ani-s \quad wave$  & 0.9 & 0.14  & 100\%    &  -  &  -   &  - \\
$two \quad s \quad waves$   & 0.7 & 0.18  & 79\%   &  0.38  &  0.07 & 21\% \\
$two \quad d \quad waves$   & 0.83 & 0.07 & 41\%     &  0.91 &  0.09 & 59\% \\
$s+d \quad waves$  & 0.58(s) & 0.13(s)  & 39\%     &  0.93(d) &  0.09(d)  & 61\% \\
 \hline \hline
\end{tabular}
\label{tab.1}
\end{table}

In Table 1, the fitting parameter $\Delta$ is the net (for isotropic s wave) or the maximum value of the superconducting gap, and it is very different when different gap functions are used. The coherence-peak energy $\Delta_{peak}$ taken directly from the spectrum in Fig.~\ref{fig2} is about 0.98$\;$meV. However, no matter what superconducting gap function we use to do the fitting, the fitting parameter $\Delta$ is always smaller than the coherence-peak energy $\Delta_{peak}$. Large zero-bias conductance of the spectrum can be attributed to large broadening parameter $\Gamma$, which broadens the spectrum and enlarges the difference between superconducting gap $\Delta$ and coherence-peak energy $\Delta_{peak}$. Considering such reasons, the wide distribution range of the coherence-peak energy $\Delta_{peak}$ shown in Fig.~\ref{fig1}(c) should be mainly caused by the difference of broadening parameter $\Gamma$ in different areas. The different broadening effect may be caused by distinct defect density in different areas of the sample.

\begin{figure}
\includegraphics[width=8.5cm]{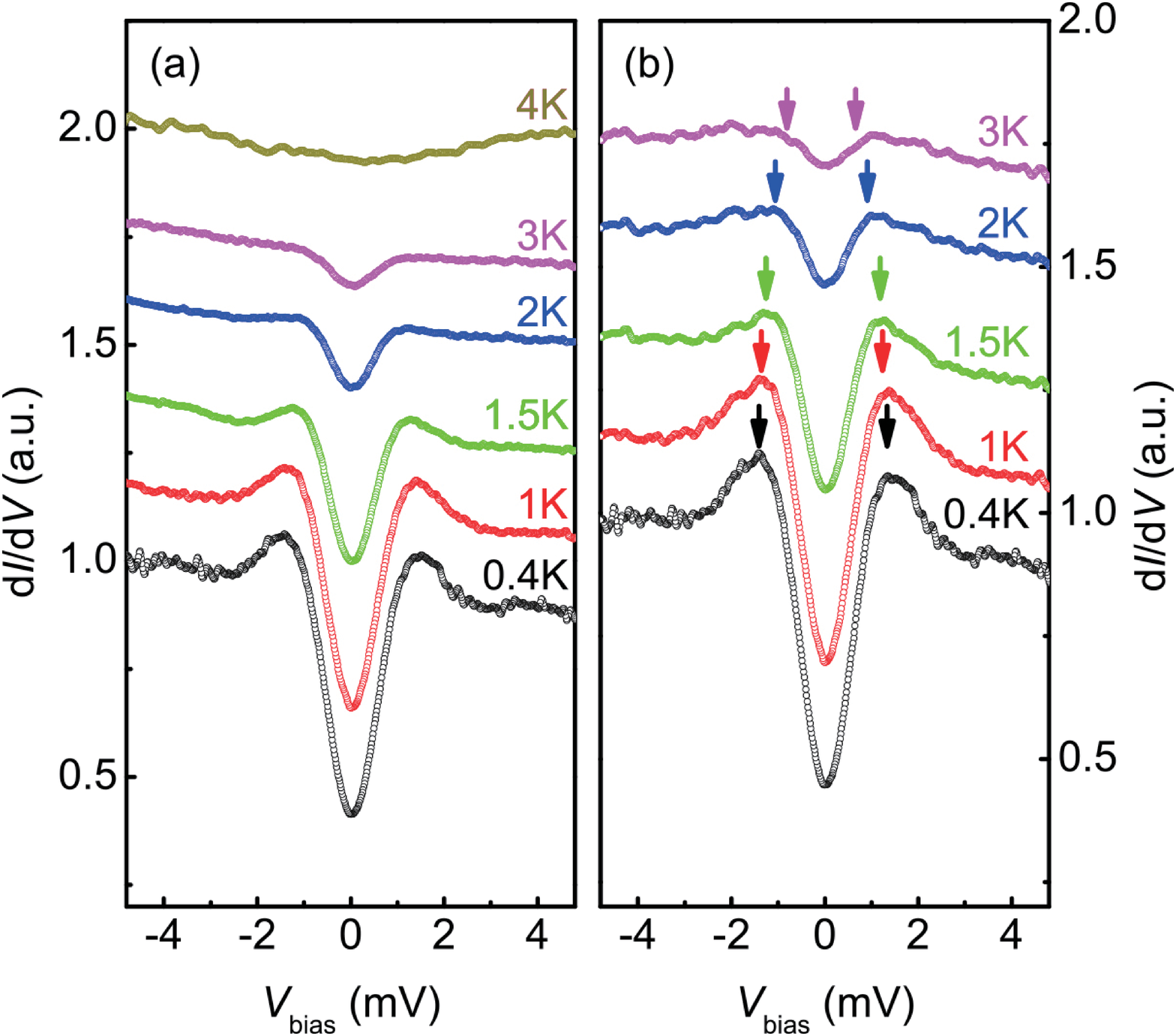}
\caption {(color online) (a) The evolution of the STS spectra measured at temperatures from 0.4 to 4$\;$K. (b) The STS spectra normalized by the one measured at 4$\;$K in the normal state. The STS spectra in (a,b) are off-set for clarity.}
\label{fig3}
\end{figure}

In Fig.~\ref{fig3}(a), we show the temperature evolution of tunneling spectra measured from 0.4 to 4$\;$K. Both of the coherence peaks and the superconducting gap are suppressed and mix together with the increase of temperature, and finally vanish at 4$\;$K. As shown in Fig.~\ref{fig3}(b), the spectra normalized by the one measured at 4$\;$K can yield well-resolved coherence peaks and superconducting gap. As guided by the colored arrows in Fig.~\ref{fig3}(b), we can see the evolution of coherence peaks with temperature.

\begin{figure}
\includegraphics[width=8.5cm]{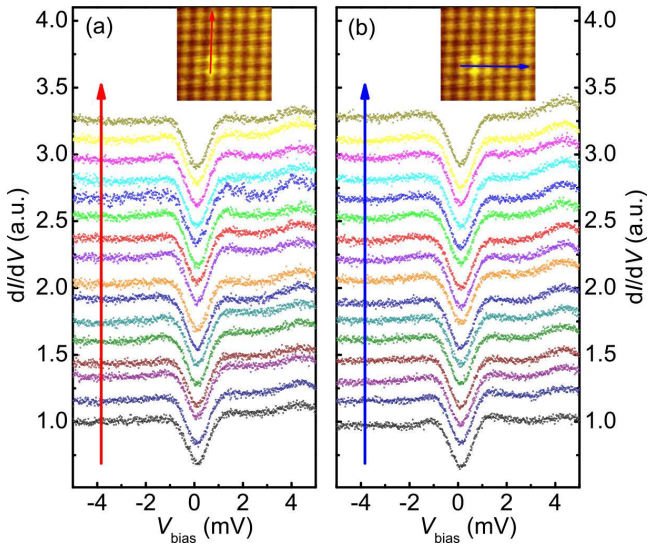}
\caption {(color online) (a, b) The spatially resolved STS spectra $dI/dV$ versus $V_{bias}$ at 0.4$\;$K. The inset of each figure shows the topographic image of the impurity. The STS spectra shown in panels (a) and (b) are measured along the trace indicated by the colored arrow (1.8$\;$nm) in the topographic image with the same spatial step. The STS spectra in panels (a) and (b) are offset for clarity.}
\label{fig4}
\end{figure}

On the image shown in Fig.~\ref{fig1}(a), we have observed some dumbbell-shaped impurities. It is necessary to investigate the influence of this kind of impurity on superconductivity. Therefore, we show a more clear topographic image around a single impurity located at the Fe site in the inset of Fig.~\ref{fig4}(a) and ~\ref{fig4}(b). We measure the STS spectra along the two high-symmetric axes of the Fe site impurity. The STS spectra seem to change weakly without resonance-state peaks when crossing the impurity site along both high-symmetric directions, which is different from those measured at Cu or Mn impurities\cite{YangH2013} in NaFe$_{1-x}$Co$_x$As. Nevertheless, the negligible effect on the superconducting gap of this Fe site impurity is similar to the one measured at Co impurity from our previous work\cite{YangH2012}. In the scenario of the $s^\pm$ pairing mechanism, both the magnetic and nonmagnetic impurities will cause in-gap states\cite{Onari2009,Bang2009,Kariyado2010,Hirschfeld2011} if the scattering potential is moderate. While nonmagnetic impurity may not induce in-gap states in the scenario of gap sign reserved pairing mechanism, or the small scattering potential when the gap changes the sign. So we can conclude that the superconducting gap either changes sign but the scattering potential on the Fe sites here are for some reason too weak, or the gap has no sign reversal at all. In addition, the coherence length value of FeS calculated from the upper critical field $H_{c2}$ is about 34$\;$nm\cite{Christopher2016}. With such a large coherence length, the impurity effect could be extended spatially. Hence, we more believe that the scattering potential of this kind of impurity may be relatively weak, or that the impurity may behave as an extended scattering center spatially, so the interpocket scattering will not be affected due to the finite size of the momentum distance. Under these circumstances, the $s^\pm$ pairing mechanism may be still relevant in the presence of such impurities, which gets further support from the analysis on another kind of defects, as shown below.

\begin{figure}
\includegraphics[width=8.4cm]{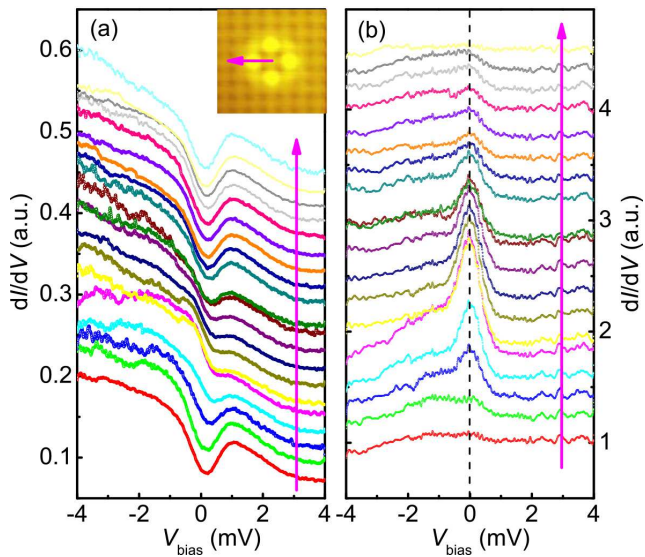}
\caption {(color online) (a) The spatially resolved STS spectra measured at the impurity located at the site of S atom. The inset in panel (a) shows the topographic image of the impurity. The purple arrow (1$\;$nm) in the topographic image indicates the measuring trace of the STS spectra. (b) The divided results of the STS spectra by the one measured at 1 nm away from the S site. The colors of the lines are the same in panel (a). The STS spectra in panel (a) and (b) are offset for clarity. The black dash line at zero-bias energy in panel (b) is a guide for eyes. One can easily see the emerging bound states when approaching the \emph{S}-site impurity.}
\label{fig5}
\end{figure}

There is another kind of impurity which turns out to be rare in our measurements. We present the topographic image of this kind of impurity in the inset of Fig.~\ref{fig5}(a); one can see that it has a fourfold symmetry shape with the center located at the position of the S atom. The tunneling spectra with bias voltage step of 0.01 mV shown in Fig.~\ref{fig5}(a) are measured along the trace indicated in the topographic image. The STS spectra exhibit an asymmetric behavior with more contribution from the occupied states, which is different from the spectra measured at other place on this sample. The reason for this type of asymmetry is unknown. A gradual evolution can be seen here with a slightly lifted height on the STS spectra near zero bias energy when approaching the S-site impurity. The STS spectra curves hardly change when the distance away from the S-site impurity is beyond 1 nm. In order to investigate how the S-site impurity influences the STS spectra, we divide the STS curves measured at different positions by the one measured 1 nm away from the center of the S-site impurity and present the results in Fig.~\ref{fig5}(b). One can clearly see that the difference of the STS spectra exhibits a bound state with the peak at zero-bias energy. To our knowledge, the  bound state generated by classical magnetic and nonmagnetic impurities should form a pair of symmetric resonance peaks with respect to zero-bias energy\cite{Yu1965,Rusinov1969,Shiba1968,ZhuJX2006}. In the unitary limit, the resonance peaks could locate deeply in the gap, so the presence of the zero-energy bound state here may suggest strong local scattering caused by the S-site impurity. This similar zero-energy bound state has been observed at the interstitial iron impurity measured on Fe(Se,Te)\cite{PanSH2015}. Nevertheless, the mechanism of this zero-energy bound state in these two systems may be different because this defect is located at the S site in FeS, not the interstitial position of Fe in Fe(Se,Te). In addition, the coherence length of FeS is about 34 nm, which may make the impurity behave as an extended scattering center spatially. Nevertheless, the experiment of impurity effect on single crystal Nb(110) observed by Yazdani \emph{et al.}\cite{Yazdani1997} has shown that the spatial effective length of the impurity bound state is about 1 nm while the coherence length of the Nb(110) sample is about 40 nm. So the effective length for the strong impurity scattering center, like the S-site impurity here, could be much smaller than the coherence length. The sharp impurity bound state on S-site impurity shown in Fig.~\ref{fig5} has very limited effective range and almost disappears in the range out of 1 nm. So we think the zero-energy bound state we get from normalizing the spectra by the one measured at 1 nm away from the S-site impurity is reliable. The emergence of the zero energy bound states would require further detailed investigation of the scattering potential of the defect and the pairing symmetry. In addition, we should mention that, the STS spectra reported in the present work show a relatively high value of zero-bias conductance. This may be partly attributed to the residual density of states arising from the impurity scattering in the possible nodal superconductor, since any inevitable disorders in the material will lead to the pair breaking and induce the ungapped density of states. This argument gets also support from the recent thermal conductivity measurements which show the residual thermal conductivity coefficient at very low temperatures\cite{LiSY}. With a large coherence length in this material, the impurity may behave as an extended scattering center spatially, which could also contribute to the finite zero-bias conductance. Another reason for the high zero bias conductivity $dI/dV$ may be the inhomogeneity problem in the early study on this superconductor. It remains to know whether this inhomogeneity problem is an intrinsic feature, such as that in underdoped cuprates, or if it is due to the sample quality. Therefore, more investigations on samples grown with other methods are highly desired in the future.

\section{SUMMARY}

In summary, we report the low-temperature scanning tunneling spectroscopy measurement on the newly discovered superconductor tetragonal FeS. The low-temperature spectrum can be fitted well by an anisotropic $s$-wave gap or $s + d$ wave gaps using the Dynes model, which suggests the existence of strong gap anisotropy, or even nodes. The corresponding gap ratio of the 2$\Delta_{max}/k_BT_c\approx4.65$ is larger than the one predicted by the BCS theory in the weak-coupling limit, suggesting strong-coupling superconductivity in this system. We have measured spatial variation of the tunneling spectra near the impurity located at the Fe site and the S site. The former impurity shows negligible effect on the superconducting gap and the STS spectra, while the latter one induces a bound state at zero-bias energy. Our discovery about the gap anisotropy and the zero-energy bound state will stimulate the investigations on the new superconductor FeS and may help to resolve the pairing mechanism in iron-based superconductors.

\begin{acknowledgments}
This work was supported by National Natural Science Foundation of China (Grants No. 11534005 and No.11190023), the Ministry of Science and Technology of China (Grants No. 2016YFA0300401, No. 2011CBA00100, and No. 2012CB821403), and PAPD.
\end{acknowledgments}

$^{\star}$ hhwen@nju.edu.cn

\end{document}